\documentclass[preprint]{aastex}

\usepackage{amssymb}
\usepackage{graphicx}
\usepackage{amsmath}

\begin{document}

\title{Is the metallicity of the progenitor of long gamma-ray bursts really
low?}

\author{Jing-Meng Hao and Ye-Fei Yuan\thanks{yfyuan@ustc.edu.cn}}
\affil{Key Laboratory for Research in Galaxies and Cosmology CAS, \\
Department of Astronomy, University of Science and Technology of China,\\
 Hefei, Anhui 230026, China}

\begin{abstract}
Observations of long gamma-ray bursts (LGRBs) offer a unique opportunity
for probing the cosmic star formation history, although whether or
not LGRB rates are biased tracers of star formation rate history is
highly debated. Based on an extensive sample of LGRBs compiled by \citet{2012ApJ...744...95R},
we analyze various models of star formation rate and the possible
effect of the evolution of cosmic metallicity under the assumption that LGRBs tend
to occur in low-metallicity galaxies. The models of star formation rate
tested in this work include empirical fits from observational data
as well as a self-consistent model calculated in the framework of the
hierarchical structure formation. Comparing with the observational
data, we find a relatively higher metallicity cut of $Z\gtrsim0.6Z_{\odot}$ for
the empirical fits and no metallicity cut for the self-consistent model. These results
imply that there is no strong metallicity preference for the
host galaxy of LGRBs, in contrast to previous work which suggest a
cut of $Z\sim0.1-0.3Z_{\odot}$, and that the inferred
dependencies of LGRBs on their host galaxy properties are strongly
related to the specific models of star formation rate. Furthermore,
a significant fraction of LGRBs occur in small dark matter halos down
to $3\times10^{8}\,\mathrm{M_{\odot}}$ can provide an alternative
explanation for the discrepancy between the star formation rate history
and LGRB rate history.
\end{abstract}


\section{Introduction}

When reionization was complete and what kind of sources should be
responsible for it still remain open questions. The optical depth of
electron scattering
constrained from the \emph{Wilkinson Microwave Anisotropy
Probe} (WMAP) infers that the universe was substantially ionized by
$z\sim10$ \citep{2011ApJS..192...18K}, which is somewhat in conflict
with the Gunn-Peterson trough in the spectra of high-redshift quasars
implying an end to reionization at $z\approx6$ \citep{2006ARA&A..44..415F}.
Models which are consistent with these constrains strongly suggest
that reionization is likely to be a much more extended process \citep{2003ApJ...591...12C,2006MNRAS.371L..55C,2007MNRAS.376..534I}.
In addition, observations of the Ly $\alpha$ forest and the high-redshift
galaxies at $z\sim6-10$ infer a photon-starved end to reionization
\citep{2007MNRAS.382..325B,2012ApJ...745..110O}. One possibility
is that small galaxies forming in the dark matter halos with masses
below $\sim10^{9}\,\mathrm{M_{\odot}}$ produce the bulk of ionizing
photons during the epoch of reionization. However, these galaxies
are too faint to be detected by the current observational facilities.
Even the \emph{James Webb Space Telescope} (JWST) will be incapable of
reaching the required sensitivity. Fortunately, long gamma-ray burst
(LGRB) observations offer a unique opportunity for probing the history
of the high-redshift star formation, unlimited by the faintness
of the host galaxy.

As a result of the collapse of massive stars \citep{1999ApJ...524..262M},
LGRBs are thought to be well suited to investigate the cosmic star
formation rate (CSFR) \citep{2001ApJ...548..522P,2002ApJ...575..111B,2008ApJ...683L...5Y,2009ApJ...705L.104K}.
Still, this is challenging, because detailed modeling is required
to connect the LGRB rate to the CSFR. In this respect, whether or
not LGRBs are biased tracers of star formation is highly debated \citep{2006MNRAS.372.1034D,2008ApJ...673L.119K}.
Earlier studies \citep[e.g.][]{2008ApJ...673L.119K} often modeled
the relation between the LGRB rate and the CSFR using a redshift
dependence quantity which is parameterized as a simple power law,
$\Psi(z)\propto(1+z)^{\beta}$, with $\beta\approx1.2$. A possible
physical explanation for such an enhancement is the cosmic
metallicity evolution, because the collapsar model for LGRBs suggests that they
can only be produced by stars with metallicity $Z\lesssim0.1Z_{\odot}$
\citep{2006ARA&A..44..507W,2006ApJ...638L..63L,2007ApJ...656L..49S}.
Observationally, \citet{2010MNRAS.405...57S} and \citet{2010AJ....140.1557L}
found that LGRBs at $z\lesssim1$ occur preferentially in relatively
low-mass, low-metallicity galaxies. However, the picture is not a
simple one: several LGRB hosts with high-metallicity have been
found \citep{2009AIPC.1133..269G,2010AJ....140.1557L,2010ApJ...712L..26L,2010ApJ...725.1337L},
which suggests that a low-metallicity cut-off is unlikely. Compiling
a large sample of 46 LGRBs over $0<z<6.3$, \citet{2009ApJ...691..182S}
found that the properties of their host galaxies are those expected
for normal star-forming galaxies. Most recently, by analyzing a sample
of 22 LGRB hosts with new radio data, \citet{2012ApJ...755...85M}
have found that the properties of LGRB population are consistent with those
of other star-forming galaxies at $z\lesssim1$, implying that LGRBs
trace a large fraction of all star formation. Hence, owing to the
limited sample size, the biases of the LGRB hosts in terms of morphology
and metallicity are far from being well understood.

In this work we investigate the effect of the evolution of cosmic metallicity
placed on the CSFR-LGRB rate connection using an extensive sample
of LGRBs compiled by \citet{2012ApJ...744...95R} together with several
CSFR models, including empirical models fitted from the observational
data as well as a self-consistent model derived from the hierarchical
formation scenario using a Press-Schechter-like formalism. This analysis
could also be used for a better estimate of the high-redshift CSFR
using the LGRB rate as the observational data. Furthermore, this work
could contribute to the study of the environments of LGRB host galaxies.
This paper is organized as follows. The CSFR and LGRB rate models
are explained in Section~2. In Section~3, we compare the predictions
of the different models with the observed cumulative redshift distribution
of LGRBs. Conclusions are presented in Section~4.

The cosmological parameters used in this paper are from the WMAP-7 results: $\Omega_{\mathrm{m}}=0.266$,
$\Omega_{\mathrm{\Lambda}}=0.734$, $\Omega_{\mathrm{b}}=0.0449$,
$h=0.71$ and $\sigma_{8}=0.801$.

\section{LGRB rate}

In order to successfully produce a LGRB with a collapsar, the progenitor
star has to be sufficiently massive to result in the formation of
a central black hole \citep{1999ApJ...524..262M}. Then the relationship
between the intrinsic LGRB rate and the black hole formation rate
can be parameterized as
\begin{equation}
\dot{n}_{\mathrm{GRB}}(z)\propto\Psi(z)\dot{n}_{\mathrm{BH}}(z),
\end{equation}
where $\dot{n}_{\mathrm{BH}}(z)$ is the black hole formation rate
and $\Psi(z)$ is the redshift-dependent LGRB formation efficiency
that can be used to model possible biases in the relation between
$\dot{n}_{\mathrm{BH}}$ and $\dot{n}_{\mathrm{GRB}}$.

\subsection{Model for $\Psi(z)$}

\citet{2008ApJ...673L.119K}
and \citet{2012ApJ...744...95R} found that $\Psi\sim\mathrm{constant}$
was inconsistent with the observational data, implying that there is an enhancement
in the LGRB rate by some mechanism at high redshift. As suggested by the
collapsar model \citep{1999ApJ...524..262M}, the most likely physical
explanation for this enhancement is the cosmic metallicity evolution,
which has been explored by many authors
\citep{2006ApJ...638L..63L,2007ApJ...656L..49S,2008MNRAS.388.1487L,2009MNRAS.400L..10W,2010ApJ...711..495B,2011MNRAS.417.3025V}.
For instance, \citet{2007ApJ...656L..49S} explored a scenario in which
LGRBs arise in metal-poor host galaxies, resulting in a metallicity
cut of $Z\lesssim0.1Z_{\odot}$. Following \citet{2006ApJ...638L..63L}
(LN), in the case where LGRBs preferentially occur in galaxies with
low-metallicity, the LGRB formation efficiency can be described by
an analytical form for the fraction of mass density belonging to metallicity
below a given threshold of $Z_{\mathrm{th}}$:
\begin{equation}
\Psi(Z_{\mathrm{th}},z)=\frac{\hat{\Gamma}[\alpha_{1}+2,(Z_{\mathrm{th}}/Z_{\odot})^{\beta}10^{0.15\beta z}]}{\Gamma(\alpha_{1}+2)},
\end{equation}
where $\hat{\Gamma}$ and $\Gamma$ are the incomplete and complete
gamma functions, $\alpha_{1}=-1.16$ is the slope in the Schechter
distribution function of galaxy stellar masses \citep{2004MNRAS.355..764P} and $\beta=2$ is the
power-law index of the galaxy mass-metallicity relation. 
It is worth stressing
that this analytical form is based on a Schechter function of
galaxy stellar masses from \citet{2004MNRAS.355..764P} and
the linear bisector fit to the mass-metallicity relation derived
by \citet{2005ApJ...635..260S}, of the form $M/M_{*}=K(Z/Z_{\odot})^{\beta}$. LN did not
address the redshift evolution of the galaxy stellar mass function, and assumed that
the average cosmic metallicity simply evolves with redshift according to $Z/Z_{\odot}\propto10^{-0.15z}$,
which is from the metallicity measurements of emission-line galaxies by \citet{2005ASSL..329..307K}.
The validity of these simplifications needs to be examined.

\begin{figure}
\begin{center}
\includegraphics[angle=0,width=0.5\textwidth]{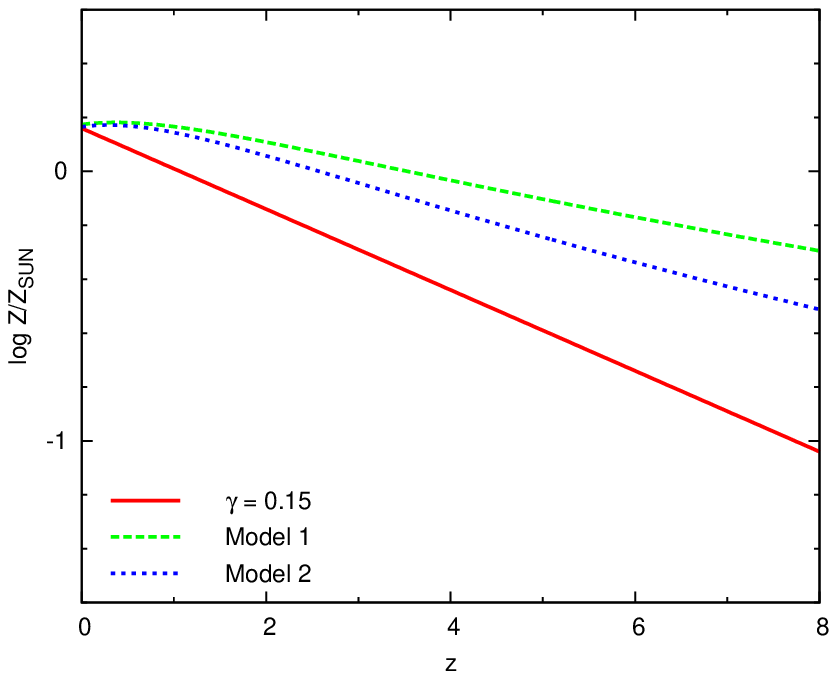}
\end{center}
\caption{Cosmic average metallicity as a function of redshift $z$.
The solid line is the evolution of the metallicity
according to $Z/Z_{\odot}\propto10^{-0.15z}$. The dashed line and dotted line are
the average metallicities calculated by adopting the redshift-dependent mass-metallicity relation of \citet{2005ApJ...635..260S},
with the non-evolving (Model 1) and evolving (Model 2) stellar mass function, respectively.}\label{fig1}
\end{figure}

Following \citet{2008MNRAS.388.1487L}, we estimate the redshift evolution
of the average metallicity below.
Given the scaling $12+\log(\mathrm{O/H})=\log(Z/Z_{\odot})+8.69$ \citep{2001ApJ...556L..63A},
the redshift-dependent mass-metallicity relation
derived by \citet{2005ApJ...635..260S} can be written as
\begin{eqnarray}
\log(Z/Z_{\odot}) & = & -16.2803+2.5315\log M\nonumber \\
 &  & -0.09649\log^{2}M\nonumber \\
 &  & +5.1733\log t_{\mathrm{H}}-0.3944\log^{2}t_{\mathrm{H}}\nonumber \\
 &  & -0.403\log t_{\mathrm{H}}\log M,\label{eq}
\end{eqnarray}
where $t_{\mathrm{H}}$ is the Hubble time in units of Gyr and $M$ is the galaxy stellar mass 
in units of $\mathrm{M_{\odot}}$.
Equation~(\ref{eq}) then can be used to calculate the average metallicity which
is defined by averaging over the stellar mass
\begin{equation}
\left\langle \frac{Z}{Z_{\odot}}\right\rangle \equiv\frac{\int_{0}^{\infty}Z(M,z)M\Phi(M)\,\mathrm{d}M}{Z_{\odot}\int_{0}^{\infty}M\Phi(M)\,\mathrm{d}M}.
\end{equation}
By adopting a redshift evolving stellar mass function from \citet{2008ApJ...680...41D},
\begin{eqnarray}
\Phi(M,z)\,\mathrm{d}M & = & \Phi_{*}\left(\frac{M}{M_{*}}\right)^{\gamma}\exp\left(-\frac{M}{M_{*}}\right)\frac{\mathrm{d}M}{M_{*}},\\
\Phi_{*} & \approx & 0.003(1+z)^{-1.07}\,\mathrm{Mpc^{-3}dex^{-1}}\nonumber \\
\log M_{*}(z) & \approx & 11.35-0.22\ln(1+z)\nonumber \\
\gamma(z) & \approx & -1.3,\nonumber
\end{eqnarray}
$\left\langle Z/Z_{\odot}\right\rangle $ is calculated and shown in Fig.~\ref{fig1},
with comparison to the measurements from \citet{2005ASSL..329..307K}. The result of
$\left\langle Z/Z_{\odot}\right\rangle $ with the non-evolving
stellar mass function from \citet{2004MNRAS.355..764P} is also shown in Fig.~\ref{fig1}.
As can be seen, the redshift evolution
of the metallicity according to $Z/Z_{\odot}\propto10^{-0.15z}$ evolves
more rapidly to lower metallicity with increasing redshift than that
of $\left\langle Z/Z_{\odot}\right\rangle $ with both the evolving
and the non-evolving stellar mass function.
This is because that the contribution to $\left\langle Z/Z_{\odot}\right\rangle $ is dominated
by galaxies with stellar masses around $M_{*}\sim10^{11}$ while faster evolution and lower metallicity
are primarily due to galaxies with smaller stellar masses \citep{2005ApJ...635..260S,2008MNRAS.388.1487L}.
However, due to the limited number of LGRBs with measured redshifts and many
uncertain biases, such as their selection effects, evolving luminosity function, 
the evolving stellar initial mass function (IMF), for our purpose, 
it is enough to adopt the analytical form of LN in this paper.

Other than LN, \citet{2012ApJ...744...95R} extended the model of
\citet{2009ApJ...702..377K} to calculate $\Psi(z)$ from the fraction
of star formation occurring below some metallicity cut. They found
that star formation occurring in galaxies with metallicity below the
value $12+\log[\mathrm{O/H}]_{\mathrm{crit}}\approx8.7$, which corresponds
to $Z\sim0.6-1.0Z_{\odot}$ depending on the adopted metallicity scale
and solar abundance value \citep{2008AJ....135.1136M}, tracks the LGRB rate with high consistency
and parameterized it as:
\begin{equation}
\Psi_{\mathrm{fit}}(z)=0.5454+(1-0.5454)\times[\mathrm{erf}(0.324675z)]^{1.45}.
\end{equation}

\begin{figure}
\begin{center}
\includegraphics[angle=0,width=0.5\textwidth]{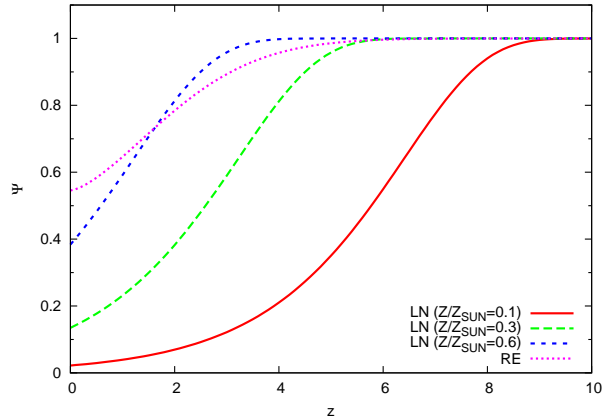}
\end{center}
\caption{Fraction of the density of the stellar mass in galaxies
with metallicities below a certain threshold of $Z/Z_{\odot}$
as a function of redshift $z$.
These results are from \citet{2006ApJ...638L..63L}(LN).
The metallicity cuts are taken to be $Z_{\mathrm{th}}=0.1$ (solid line), 0.3 (dashed line), 0.6 (short-dashed line), respectively.
A parameterized fit from \citet{2012ApJ...744...95R} (RE) is also
shown in dotted line for comparison.}\label{fig2}
\end{figure}

In Fig.~\ref{fig2}, we show a comparison of equation (2) with different
values in metallicity cut ($Z_{\mathrm{th}}=0.1-0.6Z_{\odot}$)
and the parameterized best-fit from \citet{2012ApJ...744...95R}. As can
be seen, the best-fit of \citet{2012ApJ...744...95R} is similar to the
$Z=0.6Z_{\odot}$ case of LN.

\subsection{The CSFR models}

The black hole formation rate $\dot{n}_{\mathrm{BH}}(z)$ is calculated
by
\begin{equation}
\dot{n}_{\mathrm{BH}}(t)=\int_{m_{\mathrm{BH}}}^{m_{\mathrm{up}}}\Phi(m)\dot{\rho}_{*}(t-\tau_{\mathrm{m}})\,\mathrm{d}m.
\end{equation}
where the lower limit of the integral, $m_{\mathrm{BH}}$, corresponds
to the minimum mass of a star that could collapse to a BH, which we
set to be $25\mathrm{M_{\odot}}$ \citep{2002ApJ...575..111B}, $\Phi(m)$
is the stellar IMF and $\dot{\rho}_{*}(t-\tau_{\mathrm{m}})$ represents
the CSFR at the retarded time $(t-\tau_{\mathrm{m}})$, where $\tau_{\mathrm{m}}$
is the lifetime of a star of mass $m$.

We consider that the IMF follows the \citet{1955ApJ...121..161S}
form $\Phi(m)=Am^{-2.35}$ and this function is normalized as $\int_{m_{\mathrm{inf}}}^{m_{\mathrm{up}}}Am^{-2.35}m\,\mathrm{d}m=1$,
where we take $m_{\mathrm{inf}}=0.1\,\mathrm{M_{\odot}}$ and $m_{\mathrm{up}}=140$
for lower and upper mass limits respectively.

The stellar lifetime $\tau_{\mathrm{m}}$ as a function of mass $m$
is given by the fit of \citet{1986FCPh...11....1S} and \citet{1997ApJ...487..704C}:
\begin{equation}
\log_{10}(\tau_{\mathrm{m}})=10.0-3.6\log_{10}\left(\frac{M}{\mathrm{M_{\odot}}}\right)+\left[\log_{10}\left(\frac{M}{\mathrm{M_{\odot}}}\right)\right]^{2}.
\end{equation}

For CSFR $\dot{\rho}_{*}$, there are many forms available in the
literature. \citet{2008ApJ...673L.119K,2009ApJ...705L.104K} and \citet{2012ApJ...744...95R}
adopted the piecewise-linear model of \citet{2006ApJ...651..142H},
which provides a good statistical fit to the available star formation
density data. However, it should be stressed that the empirical fit
will obviously vary depending on the functional form as well as the
observational data used. As a comparison, we also consider the model
of \citet{2001MNRAS.326..255C}, which use the parametric form:
\begin{equation}
\dot{\rho}_{*}=\frac{(a+bz)h}{1+(z/c)^{d}},
\end{equation}
where $h=0.7$, $a=0.017$, $b=0.13$, $c=3.3$ and $d=5.3$ \citep{2006ApJ...651..142H}.

In addition, we utilize a self-consistent model of \citet{2010MNRAS.401.1924P}.
In the framework of hierarchical structure formation using a Press-Schechter-like
formalism, \citet{2010MNRAS.401.1924P} obtained the CSFR by means
of solving the equations governing the total gas density taking into
account the baryon accretion rate and the lifetime of the stars formed
in the dark matter halos. We show two model predictions with different assumptions
on the threshold dark matter halo mass below which galaxy formation
is suppressed: $M_{\mathrm{min}}=3\times10^{8}\,\mathrm{M_{\odot}}$
and $M_{\mathrm{min}}=3\times10^{9}\,\mathrm{M_{\odot}}$. The lower
value assumes the star formation proceeds in dark matter halos down
to the limit of HI cooling ($T_{\mathrm{vir}}\sim2\times10^{4}\,\mathrm{K}$),
while the higher value corresponds to a fit to the observed high redshift
CSFR, which successfully reproduces the CSFR from $z=5$ to $z=8$.

\begin{figure}
\begin{center}
\includegraphics[angle=0,width=0.5\textwidth]{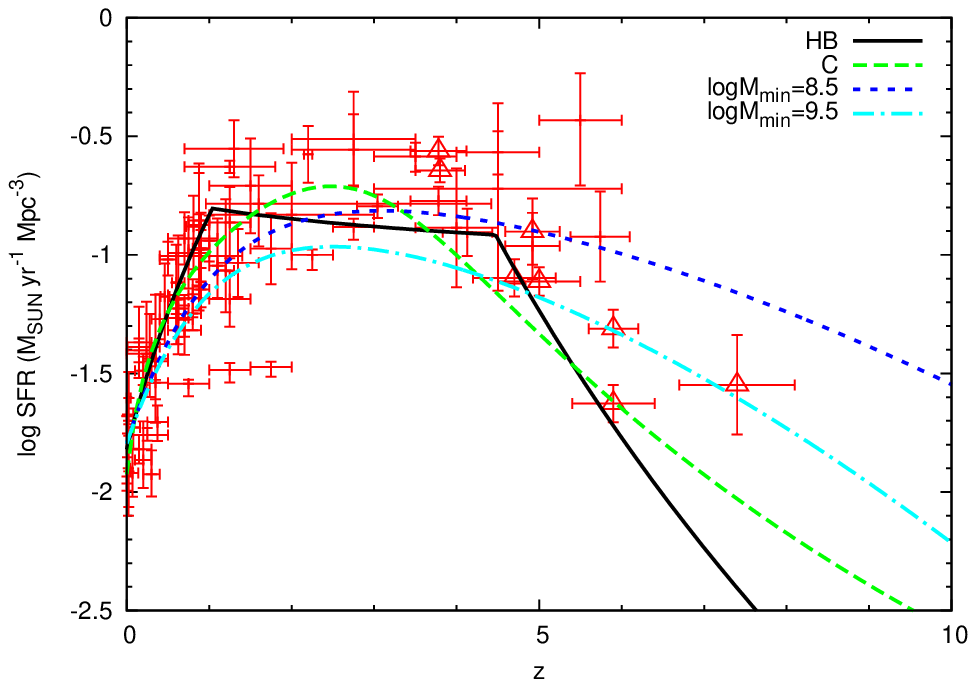}
\end{center}
\caption{Cosmic Star Formation Rate (CSFR) versus redshift $z$.
The solid line represents the empirical fit of \citet{2006ApJ...651..142H}
(HB) and the dashed line corresponds the empirical fit of \citet{2001MNRAS.326..255C} (C).
The short-dashed line and dot-dashed line represent the model of \citet{2010MNRAS.401.1924P}
with a threshold mass: $\log M_{\mathrm{min}}=8.5$ and $\log M_{\mathrm{min}}=9.5$, respectively.
The observational data is taken from \citet{2004ApJ...615..209H,2007ApJ...654.1175H}
and \citet{2008MNRAS.388.1487L}.}\label{fig3}
\end{figure}

All these different CSFRs are summarized in Fig.~\ref{fig3}, compared
to data from \citet{2004ApJ...615..209H,2007ApJ...654.1175H} and
\citet{2008MNRAS.388.1487L}. As can be seen, all of these models
are similar and have good agreement with observational data at redshift
$z<4$. At high redshifts, the \citet{2010MNRAS.401.1924P} CSFR remains
much flatter than the two empirical fits from \citet{2006ApJ...651..142H}
and \citet{2001MNRAS.326..255C}, which are already beginning to drop
exponentially.

\section{Comparison with the observational data}

In order to compare with observations, we calculate the expected cumulative
redshift distribution of LGRBs as
\begin{equation}
N(<z)=A\int_{0}^{z}\Psi(z)\dot{n}_{\mathrm{BH}}(z)\frac{\mathrm{d}V}{\mathrm{d}z}\frac{\mathrm{d}z}{1+z},
\end{equation}
where $A$ is a constant that depends on the observing time, sky coverage,
the survey flux limit and so on. $\mathrm{d}V/\mathrm{d}z$ is the
comoving volume element per unit redshift, given by
\begin{equation}
\frac{\mathrm{d}V}{\mathrm{d}z}=\frac{4\pi cd_{\mathrm{L}}^{2}}{1+z}\left|\frac{\mathrm{d}t}{\mathrm{d}z}\right|,
\end{equation}
where $d_{\mathrm{L}}$ is the luminosity distance and $\mathrm{d}t/\mathrm{d}z$
is given by \citep{2010MNRAS.401.1924P}
\begin{equation}
\frac{\mathrm{d}t}{\mathrm{d}z}=\frac{9.78h^{-1}\mathrm{Gyr}}{(1+z)\sqrt{\Omega_{\Lambda}+\Omega_{\mathrm{m}}(1+z)^{3}}}.
\end{equation}
The constant $A$ can be removed by simply normalizing the cumulative
redshift of GRBs to $N(0,z_{\mathrm{max}})$, as
\begin{equation}
N(<z|z_{\mathrm{max}})=\frac{N(0,z)}{N(0,z_{\mathrm{max}})}.
\end{equation}

Our LGRB sample is taken from \citet{2012ApJ...744...95R}, which
is consist of 162 long GRBs with measured redshifts or redshift limits.
\citet{2012ApJ...744...95R} chose the sample from \citet{2007ApJ...671..656B},
\citet{2009AJ....138.1690P}, \citet{2010ApJ...711..495B}, \citet{2011ApJS..195....2S},
\citet{2011A&A...526A..30G} and \citet{2011A&A...534A.108K}, including
only LGRBs occurring before the end of the Second \emph{Swift} BAT
GRB Catalog. To remove the influence of the \emph{Swift} threshold
owing to which low luminosity bursts could not have been seen at higher
$z$, as in \citet{2008ApJ...673L.119K} and \citet{2012ApJ...744...95R},
we use bursts only with isotropic-equivalent luminosities $L_{\mathrm{iso}}>10^{51}\mathrm{ergs\, s^{-1}}$
which is computed by
\begin{equation}
L_{\mathrm{iso}}=\frac{E_{\mathrm{iso}}}{T_{90}/(1+z)},
\end{equation}
where $E_{\mathrm{iso}}$ is the isotropic-equivalent energy and $T_{90}$
is the time interval containing 90\% of the prompt emission. This
culling leaves us 87 GRBs over $0<z<4$. For more details on the burst
sample, see \citet{2012ApJ...744...95R}.

\begin{figure}
\begin{center}
\includegraphics[angle=0,width=0.5\textwidth]{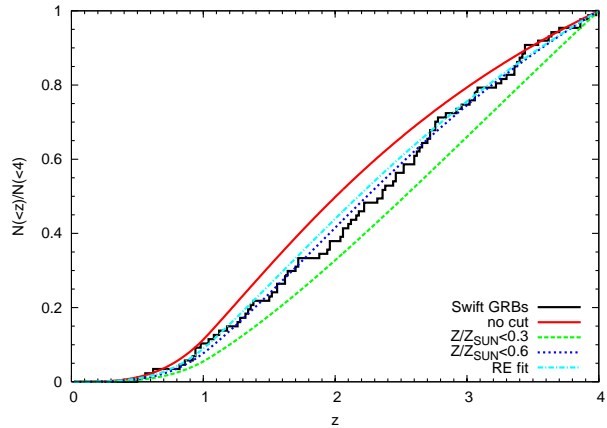}
\end{center}
\caption{Cumulative redshift distribution of LGRBs with $z<4$
and $L_{\mathrm{iso}}>10^{51}\mathrm{ergs\, s^{-1}}$.
The observational sample with 87 LGRBs are from \citet{2012ApJ...744...95R}.
The model distributions are
calculated assuming the \citet{2006ApJ...651..142H} CSFR model, for three
choices of the metallicity cuts given by Equation (2) and the parameterized best-fit from \citet{2012ApJ...744...95R}.}\label{fig4}
\end{figure}

\begin{figure}
\begin{center}
\includegraphics[angle=0,width=0.5\textwidth]{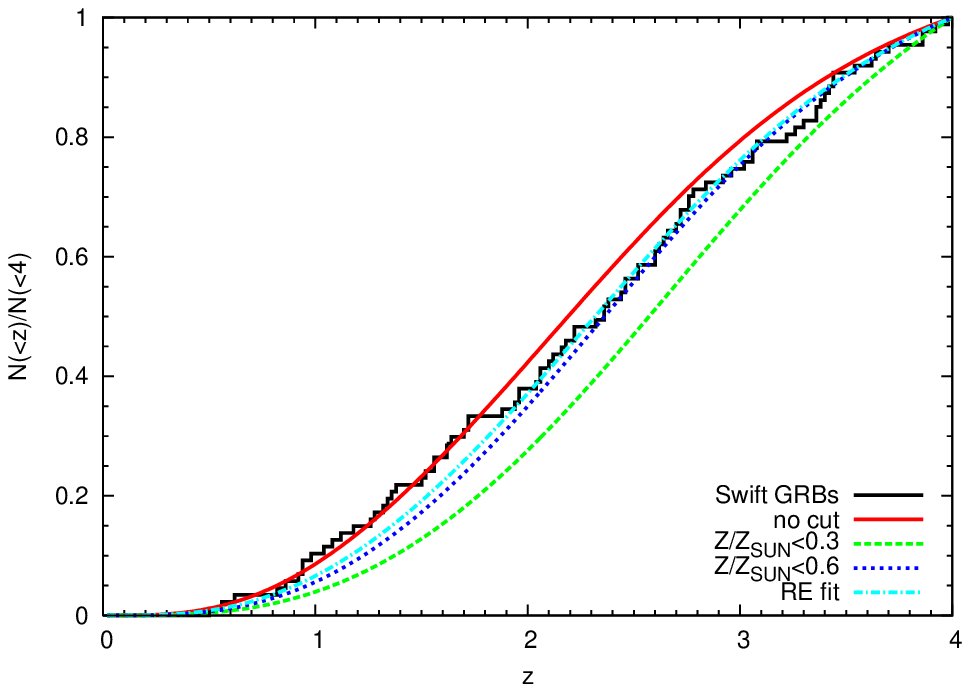}
\end{center}
\caption{Same as Fig.~\ref{fig4}, but the star formation rate is
based on the model of  \citet{2001MNRAS.326..255C} } \label{fig5}
\end{figure}

Fig.~\ref{fig4} shows the comparison between the cumulative redshift
distribution of observed LGRBs and the expectations $N(<z|z_{\mathrm{max}}=4)$
with the adoption of the \citet{2006ApJ...651..142H} CSFR model,
for three choices of the metallicity cut and the parameterized best-fit
of \citet{2012ApJ...744...95R}. We then use the one-sample Kolmogorov-Smirnov
(K-S) test to evaluate the consistency between the observed
and expected LGRB redshift distributions. In agreement with previous
studies \citep{2008ApJ...673L.119K,2012ApJ...744...95R}, the model
with no metallicity cut shows little consistency with the observations,
with $P\approx0.1$. However, in contrast to previous studies that
suggest a metallicity cut of $Z_{\mathrm{th}}\lesssim0.3Z_{\odot}$
\citep{2006ApJ...637..914W,2006ApJ...638L..63L,2007ApJ...656L..49S,2008MNRAS.388.1487L,2010MNRAS.407.1972C},
the model with a cut of $Z_{\mathrm{th}}=0.3Z_{\odot}$ shows little
consistency with the data. Only the intermediate model adopting the
value of $Z_{\mathrm{th}}=0.6Z_{\odot}$ shows high consistency with
the data, similar to the model from the best-fit of \citet{2012ApJ...744...95R}.
On the other hand, when assuming the \citet{2001MNRAS.326..255C}
model for the star formation rate, even the model with no metallicity
cut is fully consistent with the data at the probability level of
0.78 (Fig.~\ref{fig5}). The K-S test gives the probability 99\%
of a more relaxed cut of $Z_{\mathrm{th}}=0.9Z_{\odot}$. Note that
this higher cut is also more consistent with recent studies of
the LGRB host galaxies \citep{2009AIPC.1133..269G,2010AJ....140.1557L,2010ApJ...712L..26L,2012ApJ...755...85M}.
The test statistics and probability for the relevant models are summarized
in Table~\ref{tab:1}.

\begin{table}
\caption{Statistical tests of the models of CSFR for a variety of metallicity cuts\label{tab:1}}
\medskip{}
\begin{tabular}{ccc}
\tableline\tableline CSFR model & Metal cut & K-S test\tabularnewline
 & ($Z/\mathrm{Z_{\odot}}$) & D-stat, Prob\tabularnewline
\hline
HB \tablenotemark{1} & no cut & 0.1289, 0.1021\tabularnewline
HB & 0.3 & 0.1309, 0.0930\tabularnewline
HB & 0.6 & 0.0465, 0.9903\tabularnewline
HB & RE fit & 0.0705, 0.7652\tabularnewline
C \tablenotemark{2} & no cut & 0.0696, 0.7793\tabularnewline
C & 0.6 & 0.0923, 0.4305\tabularnewline
C & 0.9 & 0.0537, 0.9587\tabularnewline
C & RE fit & 0.0714, 0.7512\tabularnewline
PM \tablenotemark{3} & no cut & 0.0620, 0.9663\tabularnewline
PM & 0.3 & 0.2218, 0.0036\tabularnewline
PM & 0.6 & 0.1107, 0.4120\tabularnewline
\hline
\end{tabular}
\\
\tablenotemark{1}{\citet{2006ApJ...651..142H}}\\
\tablenotemark{2}{\citet{2001MNRAS.326..255C}}\\
\tablenotemark{3}{\citet{2010MNRAS.401.1924P}}

\end{table}

\begin{figure}
\begin{center}
\includegraphics[angle=0,width=0.5\textwidth]{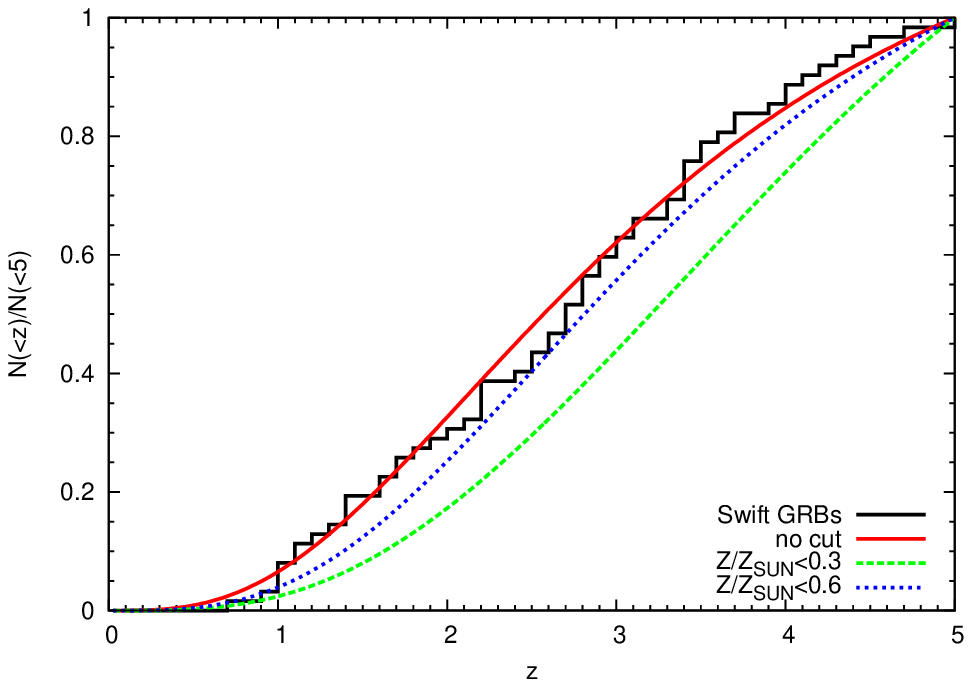}
\end{center}
\caption{Cumulative redshift distribution of LGRBs with $z<5$
and $L_{\mathrm{iso}}>3\times10^{51}\mathrm{ergs\, s^{-1}}$.
The observational sample with 62 LGRBs are from \citet{2012ApJ...744...95R},
and the theoretical distributions $N(<z|z_{\mathrm{max}}=5)$ are
based on the self-consistent star formation rate model of \citet{2010MNRAS.401.1924P}
with different metallicity cuts. The threshold mass of dark matter halo
is $\log M_{\mathrm{min}}=8.5$.} \label{fig6}
\end{figure}

\begin{figure}
\begin{center}
\includegraphics[angle=0,width=0.5\textwidth]{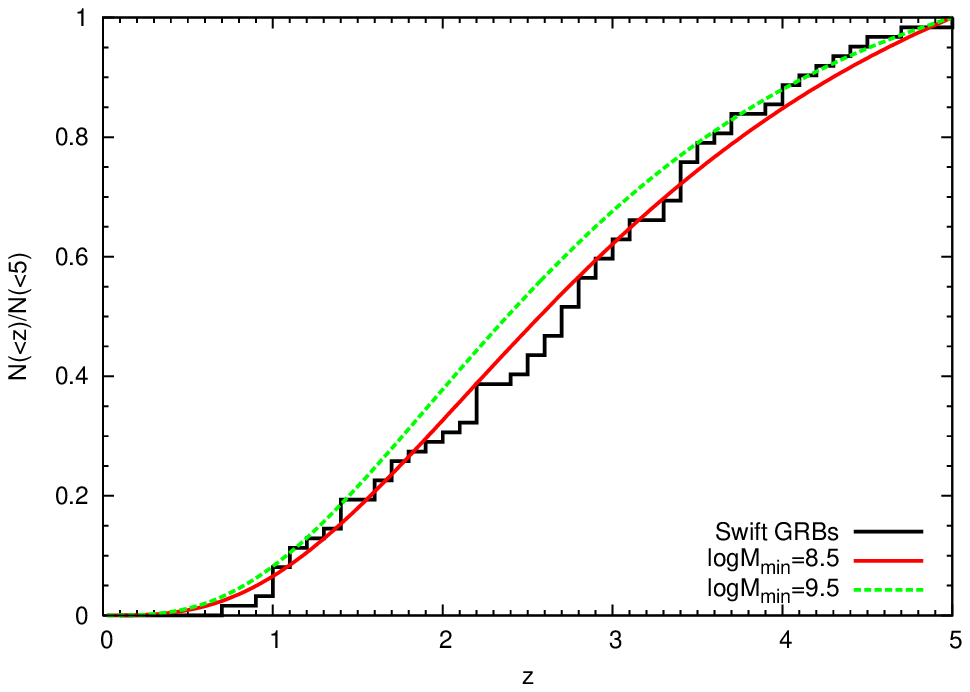}
\end{center}
\caption{Theoretical distributions of LGRBs in the
CSFR model of \citet{2010MNRAS.401.1924P}
with two threshold masses( $\log M_{\mathrm{min}}=8.5$
and $\log M_{\mathrm{min}}=9.5$).} \label{fig7}
\end{figure}

We now consider the self-consistent CSFR model of \citet{2010MNRAS.401.1924P}.
Fig.~\ref{fig6} shows a comparison with the cumulative redshift
distribution of the 62 LGRBs with $z<5$ and $L>3\times10^{51}\,\mathrm{erg\, s^{-1}}$,
normalized over the redshift range $0<z<5$. As can be seen, provided
that the star formation proceeds in dark matter halos down to the limit of
HI cooling ($T_{\mathrm{vir}}\sim2\times10^{4}\,\mathrm{K}$ and $M_{\mathrm{DM}}\sim3\times10^{8}\,\mathrm{M_{\odot}}$),
the calculated LGRB redshift distribution $N(<z|z_{\mathrm{max}}=5)$
fits the observational data very well even without considering the
extra evolution effect of metallicity ($P\approx0.96$), implying
that LGRBs are occurring in any type of galaxy. This result also implies
an alternative explanation for the CSFR-LGRB rate discrepancy, i.e.,
there is significant star formation in faint galaxies, as suggested
by \citet{2012ApJ...749L..38T}. To illustrate this, we utilize
this CSFR model to calculate the LGRB distributions for different
threshold masses of dark matter halos. The results are shown in Fig.~\ref{fig7}
and demonstrate that the LGRB redshift distribution is consistent
with a threshold halo mass of $M_{\mathrm{min}}=3\times10^{8}\,\mathrm{M_{\odot}}$
at 96\% level (and $M_{\mathrm{min}}=3\times10^{9}\,\mathrm{M_{\odot}}$
at 39\% level). This is also in agreement with what is found by \citet{2011ApJ...729...99M},
in which the minimum mass halo capable of hosting galaxies is suggested 
to be around $2.5\times10^{9}\,\mathrm{M_{\odot}}$.

\section{Conclusion}

The association of LGRBs with the death of massive stars has presented
a unique opportunity for probing the history of star formation at high redshift.
In this case, the manner in which how the LGRB rate traces the
CSFR should to be known. In this work, we have investigated the idea
that LGRBs as biased tracers of the CSFR occur preferentially in galaxies
with low-metallicity. We have tested various CSFR models together
with the metallicity considerations of \citet{2006ApJ...638L..63L}
using the constraints from newly discovered bursts.

Comparing with the cumulative redshift distribution of luminous
($L_{\mathrm{iso}}>10^{51}\mathrm{ergs\, s^{-1}}$)
\emph{Swift} LGRBs compiled by \citet{2012ApJ...744...95R} over
$0<z<4$, we find a relatively higher metallicity cut of $Z_{\mathrm{th}}=0.6-0.9Z_{\odot}$
for both star formation rate models of \citet{2006ApJ...651..142H}
and \citet{2001MNRAS.326..255C}, in contrast to previous studies
which suggest a strong metallicity cut of $\sim0.1-0.3Z_{\odot}$
\citep{2007ApJ...656L..49S,2010MNRAS.407.1972C,2011MNRAS.417.3025V}.
Especially when considering a self-consistent star formation model
of \citet{2010MNRAS.401.1924P} that takes into account a hierarchical
structure formation scenario using a Press-Schechter-like formalism,
the calculated expectations show strong consistency with the observational
data over $0<z<5$, requiring no metallicity cut at all. These results
imply that LGRBs trace a large fraction of all star formation with
no preference on the properties of their host galaxies, and are therefore
less biased indicators than previously thought, which is consistent
with recent studies on LGRB hosts \citep{2012ApJ...755...85M,2012A&A...539A.113E}.
Therefore, we conclude that LGRBs populate all types of star-forming
galaxies, with no strong metallicity preference. Using the self-consistent
CSFR model, we also find that the scenario that a significant fraction of LGRBs occur
in small dark matter halos down to $3\times10^{8}\,\mathrm{M_{\odot}}$
can provide an alternative explanation for the discrepancy between
the CSFR history and LGRB rate history. Our results also show that
the inferred dependencies of LGRBs on their host galaxy properties
are strongly related to the specific CSFR model one adopts, suggesting that
detailed observations of individual LGRB host galaxies are essential
to provide a better understanding of the metallicity cut for LGRB
production. If numbers of similar observations are confirmed, it could mean that
the key role that metallicity plays in the production of LGRBs, which
is suggested by the traditional collapsar model, needs reconsideration
in future studies or it may need alternative progenitor pathways that
do not necessarily require a low-metallicity environment.

\acknowledgements
We thank the anonymous referee for her/his useful suggestions which 
are helpful for improving the manuscript.
This work is partially supported by
National Basic Research Program of China (2009CB824800, 2012CB821800),
the National Natural Science Foundation (11073020, 11133005, 11233003),
and the Fundamental Research Funds for the Central Universities (WK2030220004).


\end{document}